\documentclass[sigconf, table]{acmart}

\renewcommand\footnotetextcopyrightpermission[1]{}

\acmConference[EASE 2025]{The 29th International Conference on Evaluation and Assessment in Software Engineering}{17–20 June, 2025}{Istanbul, Türkiye}

\AtBeginDocument{%
  }

\usepackage{subcaption}
\usepackage{textcomp}
\usepackage{multirow}
\usepackage{caption}
\usepackage{booktabs}
\usepackage{array}

\begin{document}

\title{Unveiling Ruby: Insights from Stack Overflow and Developer Survey}

\author{Nikta Akbarpour, Ahmad Saleem Mirza, Erfan Raoofian, Fatemeh Fard, Gema Rodríguez-Pérez}
\affiliation{%
  \institution{University of British Columbia, Department of Computer Science}
  \city{Kelowna}
  \state{BC}
  \country{Canada}
}
\email{{niktakbr; ahmadsm1}@student.ubc.ca, {erfan.raoofian; fatemeh.fard; gema.rodriguezperez}@ubc.ca}

\begin{abstract}
Ruby is a widely used open-source programming language, valued for its simplicity, especially in web development. Despite its popularity, with over one million users on GitHub, little is known about the issues faced by Ruby developers. 
This study aims to investigate the key topics, trends, and difficulties faced by Ruby developers by analyzing over $498,000$ Ruby-related questions on Stack Overflow (SO), followed by a survey of  $154$ Ruby developers. We employed BERTopic modeling and manual analysis to develop a taxonomy of $35$ topics, grouped into six main categories. Our findings reveal that Web Application Development is the most commonly discussed category, while Ruby Gem Installation and Configuration Issues emerged as the most challenging topic. Analysis of trends on SO showed a steady decline. A survey of $154$ Ruby developers demonstrated that $31.6\%$ of the participants find the Core Ruby Concepts category particularly difficult, while Application Quality and Security is found to be difficult for over $40\%$ of experienced developers. Notably, a comparison between survey responses and SO metrics highlights a misalignment, suggesting that perceived difficulty and objective indicators from SO differ; emphasizing the need for improved metrics to capture developer challenges better. Our study provides insights about the challenges Ruby developers face and strong implications for researchers. 
\end{abstract}



\maketitle

\section{Introduction}
Ruby is an open-source programming language widely used for a variety of applications, including web development, scripting, automation, and game design \cite{chandrancorrelative}. 
Its dynamic and object-oriented features make it a preferred choice for developers solving complex problems \cite{flanagan2008ruby}.
Unlike other scripting languages like JavaScript, Python, and PHP, Ruby strikes a unique balance between functional and imperative programming. Its design, which treats every component as an object, simplifies code structure and promotes clarity. Features such as blocks and mixins enable developers to create reusable and modular code, setting Ruby apart from its peers. Additionally, Ruby's high customizability allows developers to modify core components as needed \cite{rubyAbout}.
The language's adoption is significant, with over $2$ million projects on GitHub utilizing Ruby across multiple domains. Major companies, such as Airbnb, GitHub, and Shopify, rely on Ruby to build and maintain their platforms, underscoring its relevance in large-scale, real-world applications \cite{kaleba2022you}.
The ongoing importance of Ruby within the software development community is further illustrated by more than $228,000$ questions \textit{tagged} ``ruby'' on Stack Overflow (SO), highlighting active engagement among developers.
Despite its many strengths, Ruby presents challenges that can affect developers' effectiveness and productivity. 
Notably, there is a lack of research specifically focused on the obstacles faced by Ruby developers. This gap in understanding may hinder the language's growth and evolution, particularly as newer and more efficient programming languages continue to emerge. Addressing these challenges is crucial for maintaining Ruby's competitiveness and relevance in the rapidly changing landscape of programming.

While numerous studies have examined the challenges developers face across various domains and programming languages, using data from SO \cite{zhang2019empirical, islam2019comprehensive, van2020formal, yang2022developers}, a significant gap remains in understanding the particular issues experienced by Ruby developers.
The limited academic attention devoted to Ruby further highlights the need for focused investigation into its unique challenges \cite{RubyBib}. 
This study aims to fill this gap by \textit{analyzing Ruby-related questions from SO to identify the key topics and challenges Ruby developers encounter followed by a survey from Ruby developers to indicate their perceptions of the SO findings}.
We address the following questions:

\textbf{RQ1: What topics are commonly raised by Ruby developers through SO questions?}
By collecting $498,719$ Ruby-related questions from SO and applying the BERTopic modeling technique, we derive key topics and create a taxonomy using a combination of GPT-4o and human analysis \cite{grootendorst2022bertopic, openai2023gpt4}.

\textbf{RQ2: What are the popularity, difficulty, and evolving trends of Ruby topics in SO posts?}
We investigate which topics from our taxonomy are the most popular and difficult and analyze how interest in these topics has changed over time. We use five specific metrics to evaluate the popularity and difficulty of each topic while examining trends separately to capture shifts across Ruby topics.

\textbf{RQ3: To what extent do Ruby developers' perceptions of topic difficulty align with the findings from our SO analysis?}
We complement our findings with a survey of Ruby developers to validate our results.
We build statistical tests and models to investigate whether Ruby developers’ perceptions of topic difficulties align with our SO analysis.

\textbf{RQ4: Which topics do Ruby developers find most time-consuming to resolve?}
We explore whether developers’ experiences in resolving issues reflect the time metrics derived from SO data.
Using statistical tests, we compare developers' responses to the derived time from SO.

Our methodology integrates BERTopic for topic modeling, large language models (LLMs) for prompting, a human-in-the-loop approach for taxonomy construction, and a developer survey to gather subjective insights. While each method is individually reasonable, the novelty lies in their combination.

We identified a hierarchical taxonomy of $35$ topics, organized into six main categories of Ruby challenges and a decline in trends of questions on SO. 
Our study also uncovered notable discrepancies between objective metrics derived from SO data and the subjective difficulty ratings provided by $154$ developers in the survey, highlighting the limitations of traditional SO metrics in capturing the full complexity of issues faced by developers. This suggests the need for more nuanced metrics that better reflect the true developer experience.

This study offers valuable insights into the main challenges that Ruby developers face, paving the path for library contributors and tool developers to create resources that effectively address these pain points and ensure improvements in the Ruby ecosystem that align with developer needs.
We open-source our collected dataset and codes to enable the reproducibility of the results.\footnote{\url{https://github.com/niktaakbarpour/Unveiling-Ruby}}

\section{Related Works}

\textbf{SO Studies.}
Research on developer challenges in software development spans a wide range of areas, revealing common issues that developers encounter. 
Research has highlighted issues encountered by big data developers \cite{bagherzadeh2019going}, security challenges \cite{yang2016security}, Docker-related concerns \cite{haque2020challenges}, and chatbot development difficulties \cite{abdellatif2020challenges}. Other studies have explored developer engagement in specific programming languages such as Swift, Go, and Rust \cite{barua2014developers, chakraborty2021developers}. Others explore SO to understand challenges in fields like web development \cite{bajaj2014mining}, mobile development \cite{rosen2016mobile}, privacy issues \cite{tahaei2020understanding}, and the engagement of developers in machine learning topics within software engineering \cite{bangash2019developers}.
Other explored topics are issues in software documentation \cite{aghajani2019software}, deployment challenges in deep learning \cite{chen2020comprehensive}, deployment issues in mobile deep learning apps \cite{chen2021empirical}, defects in TensorFlow applications \cite{zhang2018empirical}, study Rust’s usability \cite{coblenz2023multimodal}, GPU programming difficulties \cite{yang2023understanding}, explore Rust’s safety rules \cite{zhu2022learning} and Rust's use in embedded systems \cite{sharma2023rust}. 
Several studies used topic modeling methods such as Latent Dirichlet Allocation (LDA) \cite{blei2003latent, bagherzadeh2019going, haque2020challenges, chakraborty2021developers,tahaei2020understanding}. Other works conducted manual analysis \cite{aghajani2019software, chen2020comprehensive, chen2021empirical, zhang2018empirical}, a combination of automated and manual methods \cite{coblenz2023multimodal, yang2023understanding}, or mixed-method approaches \cite{zhu2022learning, sharma2023rust} to find developers' challenges. 

\textbf{BERTopic.}
Other than LDA, semi-supervised topic modeling and LLMs have been applied to align documentation with developer queries \cite{cogo2023assessing}, categorize SO questions \cite{beyer2018automatically}, and understand user intents \cite{shah2023using}.
However, recent works show that BERTopic, a newer topic modeling method, outperforms previous techniques in clustering, topic separation, semantic clarity, and extracting meaningful insights from text \cite{lee1999learning, angelov2020top2vec, gan2023experimental, egger2022topic, jung2024expansive}. This model has been effectively used to analyze challenges faced by asset management tool users \cite{zhao2024empirical}, to study developer migration and topic trends within the Jira infrastructure \cite{diamantopoulos2023semantically, gu2023self}, 
and to analyze millions of SO posts about Android APIs \cite{naghshzan2023enhancing}.
Given BERTopic’s strengths, we adopt it in our study to overcome the limitations of traditional topic modeling techniques.

\textbf{Studies on Ruby.}
Ruby has been the subject of various studies in software engineering, with research examining its community dynamics, evolution, and popularity over time.  
Ruby's development history was analyzed through hosting platforms like RubyForge and RubyGems, highlighting changes in project growth and developer activity \cite{squire2018data}. The security vulnerability of RubyGems dependency networks was studied in \cite{zerouali2022impact}. 
Others have explored Ruby developers' dynamics \cite{orlowska2021programming, tambad2020analyzing}.
Longitudinal SO studies from $2008$ to $2020$ show that Ruby’s share of user engagement peaked at around $6\%$ in $2012$ but declined to $2\%$ by $2020$, reflecting a shift in developer interest toward other web frameworks \cite{moutidis2021community}. 
However, studies on developer communities show that Ruby fosters an active user base on GitHub and SO though its popularity has declined since peaking in $2008$ \cite{orlowska2021programming, tambad2020analyzing}. Ruby remains a preferred language in education, valued by students for its readability and productivity in web development contexts \cite{islam2024comparative}. Additionally, GitHub management studies indicate Ruby's continued presence on the platform, with frequent use of tools like milestones to facilitate project organization \cite{zhang2020github}.
Ruby has over {one million} users {on GitHub} and its web application framework, Ruby on Rails, is used in over $400,000$ websites\footnote{\url{https://www.bacancytechnology.com/blog/ruby-on-rails-statistics-and-facts}}.
These works demonstrate that though SO surveys show a decline in using Ruby, there is still an active community of developers and therefore, studying the main challenges faced by developers remains an open research gap. 

\textbf{Novelty of our work.}
While the current literature covers developers' challenges in other programming languages and provides insights into Ruby's popularity, community engagement, and collaborative practices, it does not identify the specific challenges faced by Ruby developers. Our study fills this gap by analyzing Ruby-related SO questions and conducting a targeted survey of Ruby developers to identify key areas of difficulty and evolving interests.

\section{Methodology}

\begin{figure*}[htbp]
\includegraphics[width=\textwidth]{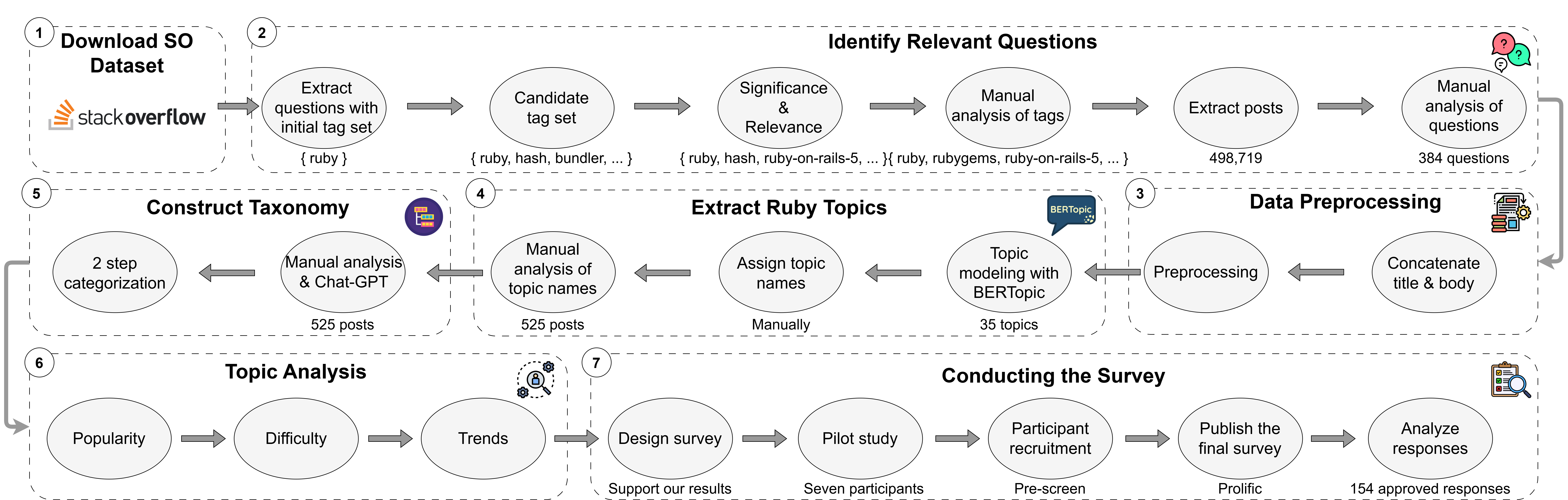}
\centering
\caption{An overview of the methodology.}
\label{fig:methodology_overview}
\end{figure*}

Figure \ref{fig:methodology_overview} shows our methodology. First, we collected the SO dataset (step \textcircled{1}), identified the relevant questions, and preprocessed the data (steps \textcircled{2}-\textcircled{3}). Then, 
we extracted Ruby topics of this dataset (step \textcircled{4}) and constructed a taxonomy of the Ruby challenges (step \textcircled{5}). Finally, we analyzed the topics in our taxonomy (step \textcircled{6}) and conducted a survey with Ruby developers to understand if we can support our experimental study results (step \textcircled{7}). The details of each step are found below.

\textbf{Step 1: Download Stack Overflow dataset.}
We acquired the SO dataset from the Stack Exchange Data Dump \cite{stackexchange2021}. Each post in the dataset includes various metadata, such as a unique identifier, creation date, title, body, and score. 
Posts are further categorized with one to five tags by SO, indicating their topics or subject matter. For questions, contributors can mark one of the answers as ``accepted,'' indicating that it resolves the issue posed.

The SO dataset we employed, denoted as $D$, consists of posts spanning from \texttt{July 31/ 2008} to \texttt{March 10/ 2024} (the time this study was conducted).

We selected SO as our primary data source due to its rich metadata, which is crucial for analyzing topics (step \textcircled{6}). Key data such as view counts, time to receive an accepted answer, and question scores are not available on platforms like GitHub.

\textbf{Step 2: Identify relevant questions.}
We employed a tagging-based approach similar to that in \cite{yang2023understanding, haque2020challenges, tahaei2020understanding}. 
First, we extracted all questions tagged with ``ruby,'' creating an initial candidate set of questions, denoted as $C$, comprising $245,467$ questions.
Next, we collected all the tags associated with questions in $C$ to form a candidate tag set, $T$, containing $10,250$ unique tags. We then refined $T$ by selecting only the tags that were significantly relevant to Ruby. This refinement process was guided by two heuristics, commonly used in prior studies \cite{yang2023understanding, ahmed2018concurrency, yang2016security}:

\begin{equation}
    \alpha (\text{relevance}) = \frac{{\text{\# questions with a specific tag } t \text{ in }} \textit{C}}{{\text{\# questions with tag } t \text{ in }} \textit{D}}
    \label{eq:alpha}
\end{equation}

\begin{equation}
    \beta (\text{significance}) = \frac{{\text{\# questions with a specific tag } t \text{ in }} \textit{C}}{{\text{\# questions} \text{ in }} \textit{C}}
    \label{eq:beta}
\end{equation}

$\alpha$ (Equation \ref{eq:alpha}) measures how relevant is a specific tag in $T$ to Ruby and $\beta$ (Equation \ref{eq:beta}) measures how significance is a specific tag in $T$. Tags meeting or exceeding defined thresholds for both heuristics were considered significantly relevant to Ruby. We applied thresholds from previous studies \cite{yang2023understanding}, setting six thresholds for $\alpha$ = \{$0.05$, $0.1$, $0.15$, $0.2$, $0.25$, $0.3$\} and six for $\beta$ = \{$0.005$, $0.01$, $0.015$, $0.02$, $0.025$, $0.03$\}.
With $7$ years of software development experience, the first author evaluated all $36$ tag configurations by manually inspecting five random questions per tag to assess their relevance to Ruby programming.
The best configuration we selected is $\alpha$ = $0.1$ and $\beta$ = $0.01$, aligning with the findings of earlier research \cite{yang2023understanding}. This refinement yielded the following tag set, $T$ = \textit{\{ruby, devise, ruby-on-rails-5, rspec, rubygems, sinatra, activerecord, ruby-on-rails-3, nokogiri, rvm, ruby-on-rails, ruby-on-rails-4, hash\}}.

However, after conducting an additional round of manual inspection, in which one author reviewed the tags and examined five questions associated with each tag, it became clear that some relevant tags were missing while others, such as \textit{hash}, were less relevant to the Ruby context. We made additional adjustments, resulting in the final refined tag set $T$ to include \textit{ruby, devise, ruby-on-rails-5, rspec, rubygems, sinatra, activerecord, ruby-on-rails-3, nokogiri, rvm, ruby-on-rails, ruby-on-rails-4, mongoid, erb, ruby-on-rails-3.2, watir, jruby, rails-activerecord, ruby-on-rails-3.1}. Using these tags, we identified the final set of Ruby-related questions, $F$, by filtering the original dataset $C$ for posts containing at least one of the selected tags.

This process yielded $498,719$ questions in $F$, which formed the basis for our subsequent analysis.

\textbf{Validation of the dataset.}
To ensure the relevance of the questions in dataset $F$ to Ruby programming, we implemented a validation procedure. First, we randomly sampled $384$ questions from $F$, which provided a $95\%$ confidence level with a $5\%$ margin of error. Each question's title and body were then reviewed in detail by two authors, {each} with seven years of development {experience}. They independently labeled each question as ``TRUE'' if it was related to Ruby and ``FALSE" if it was not. 
The inter-rater agreement between the two reviewers, measured by Cohen’s Kappa \cite{mchugh2012interrater}, was $0.86$, indicating almost perfect agreement.  After resolving any disagreements, we concluded that $98.43\%$ of the questions in $F$ were indeed relevant to Ruby.

\textbf{Step 3: Data preprocessing.}
We preprocessed our dataset in multiple stages. First, we removed posts 
with titles and bodies under $50$ characters ($4,398$ posts), as manual inspection showed they lacked meaningful context. For the rest, we concatenated titles and bodies since our manual analysis showed both contained essential keywords. To create embeddings we removed punctuation marks and stop words from NLTK list \cite{bird2009natural}. For creating the top $10$ words for each topic, we further removed escape characters, HTML tags, links, code, numbers, and applied lowercasing as used in previous studies \cite{haque2020challenges, yang2023understanding, bagherzadeh2019going, ahmed2018concurrency, zhao2024empirical}.

\textbf{Step 4: Extract Ruby topics.}
We utilized BERTopic \cite{grootendorst2022bertopic} for extracting topics related to Ruby programming as it has shown promising results in previous studies \cite{diamantopoulos2023semantically, gu2023self, tao2023code, zhao2024empirical} and has outperformed other topic modeling methods \cite{grootendorst2022bertopic, gan2023experimental, egger2022topic}. 

To use BERTopic, we generated text embeddings for clustering using e5-mistral-7b-instruct model \cite{wang2023improving}, which has demonstrated top performance in clustering-related tasks.
Our choice of the e5-mistral-7b-instruct model was based on several technical advantages over QA-specific models: First, the e5-mistral-7b-instruct model contains $7.11$ billion parameters, compared to multi-qa-MiniLM-L6-dot-v1 ($109$M parameters), multi-qa-distilbert-dot-v1 ($66.4$M parameters), and multi-qa-mpnet-base-dot-v1 ($22.7$M parameters). Second, e5-mistral-7b-instruct is trained on QA datasets like TriviaQA and MS-MARCO. Furthermore, based on \cite{cao2024recent}, the e5-mistral-7b-instruct model ranked $4th$ among the best-performing text embedding methods on the MTEB English benchmark, with an average score of $66.63$. Lastly, on the MTEP leaderboard \cite{muennighoff2022mteb}, the e5-mistral-7b-instruct model achieved a v-measure score of $50.26$ for clustering tasks.
Instead of using the default HDBSCAN algorithm \cite{mcinnes2017hdbscan} as in BERTopic, we employed K-Means clustering \cite{fix1951discriminatory}. This decision was driven by the fact that K-Means yielded more accurate results for our dataset in comparison to HDBSCAN.
To determine the optimal number of topics (i.e., $K$) for our modeling process, we employed the elbow method \cite{thorndike1953belongs}. 
To characterize each cluster, we identified the top $10$ words that best represent the respective topics. This process began with the application of CountVectorizer \cite{fabian2011scikit}, followed by the TF-IDF (Term Frequency-Inverse Document Frequency) transformation.
Initially, the elbow method suggested setting $K$ to $40$. However, upon conducting the topic modeling and manually reviewing the top $10$ keywords, we observed overlaps among some topics. Consequently, several topics were merged, resulting in a final set of $35$ distinct topics.
The merging criteria were primarily based on keyword overlap between topics, supplemented by a detailed manual examination of the keywords and representative post samples within each cluster. To ensure consistency and validity in our merging decisions, two authors independently reviewed the topics.

We conducted thorough experimentation with various hyperparameters and algorithms within BERTopic. After obtaining results for each configuration, the first author manually reviewed the top $10$ questions within each topic, ranked by their probability of belonging to each cluster. Questions were labeled as \texttt{TRUE} if both the title and body aligned with the top $10$ keywords for that topic, and \texttt{FALSE} otherwise. We also accounted for the number of questions that were identified as outliers, meaning they did not fit into any established topic. This manual analysis, conducted after each configuration, ultimately led to the selection of K-Means with the elbow method as the final approach for our topic modeling.

To assign names to each topic, one of the authors reviewed the top $10$ words generated by BERTopic, along with the titles and bodies of seven representative posts per topic (i.e., $245$ posts). Based on this analysis the author proposed a topic name. A second author then independently verified these topic names by reviewing the same set of titles and bodies, after which a consensus on the final topic names was reached.
Due to the high number of topics and the rigorous manual effort in choosing the best clustering algorithm and hyperparameter setting initially, we opted to evaluate seven questions per topic, as it showed to be enough for the manual evaluation at this stage.

We validated the topic names using $15$ additional random posts per topic, totaling $525$ posts, and two authors reviewed the titles and bodies of these questions, labeling each as \texttt{TRUE} if it belonged to the assigned topic and \texttt{FALSE} if it did not. The inter-rater agreement between the two authors, measured using Cohen’s Kappa, was $0.98$, indicating the highest level of agreement.

\textbf{Step 5: Construct taxonomy of challenges.}

After conducting topic modeling, we needed to develop a topic taxonomy to provide a clearer understanding of the hierarchy of related challenges. This taxonomy would also support the subsequent analysis. Our goal was to create a three-level taxonomy consisting of topics, middle categories, and main categories. At the lowest level, we started with the $35$ topics generated in step \textcircled{4}. The second level, referred to as the ``middle categories,'' groups related topics into broader groups. The third and highest level consists of the ``main categories,'' which are made by grouping middle categories.

To construct the taxonomy, we used a combination of GPT-4o and a human-in-the-loop approach, a methodology used in prior studies \cite{shah2023using}. The process began by providing GPT-4o with the titles and bodies of $15$ randomly selected questions from each of the $35$ topics, as well as the topic names. GPT-4o was instructed to generate a one-level taxonomy, aiming to create $15$ to $20$ middle categories based on this information and assign names to each of them.
Next, two authors manually reviewed the titles and bodies of the $525$ questions (i.e., $15$ per topic), as well as the topic names. After closely examining the taxonomy generated by GPT-4o, the authors refined the groupings, ultimately producing $18$ middle categories.
In the final step, we repeated this process to develop the highest level of the taxonomy. GPT-4o was asked to generate $5$ to $10$ higher-level groupings based on $18$ middle categories and $15$ randomly selected questions from each of them (i.e., $270$ posts), which would serve as the main categories. As a result, \textit{six} main categories were created.
To validate the categories, we sampled $525$ posts—exceeding the $384$ needed for $95\%$ confidence with a $5\%$ margin of error. Two authors independently labeled them into six categories, achieving a Cohen’s Kappa of $0.98$, indicating near-perfect agreement. This validation ensured that all Ruby-related questions could fit within our six-category taxonomy.

This approach aligns with methodologies recently adopted in Software Engineering studies that leverage advanced language models for taxonomy development and topic modeling. For instance, studies such as \cite{wang2023prompting} propose aggregating sentence-level topics using LLMs to extract topics across varying datasets. Similarly, an LLM-driven taxonomy methodology for user intent analysis in log data demonstrated how LLMs, combined with human validation, effectively generate and validate meaningful categories for analysis \cite{shah2023using}. By employing BERTopic and validation techniques, our methodology follows a similar trajectory, ensuring that our taxonomy is precise while benefiting from advanced embedding techniques and validation best practices established in these studies.

\textbf{Step 6: Topic analysis}. \label{step6}
To assess the popularity and difficulty of topics, we adopted methods from prior research \cite{yang2023understanding, haque2020challenges}. 
\textit{Popularity} is evaluated using two metrics: the average number of views for questions within each topic and the average score of questions in each topic.
The average number of views shows how many times that question was viewed by people on SO. When a question is useful to the community, users can upvote that question and otherwise, downvote it. The score of each question is calculated by summing the number of upvotes minus the number of downvotes. So, if a question has a high average score it means that the question was very useful to the community. Following previous studies \cite{yang2023understanding}, our main metric for popularity is the average number of views.
The \textit{difficulty} is assessed through three distinct metrics that were used in previous studies \cite{yang2023understanding, haque2020challenges}. 
First, we considered the percentage of questions within each topic that do not have an accepted answer. Second, we evaluated the percentage of questions that remain unanswered. Third, we measured the median time required for questions in each topic to receive an accepted answer.
Following previous studies \cite{yang2023understanding}, our main metric for understanding difficulty is the percentage of questions without an accepted answer. The author of the question can mark one of the answers as an accepted answer. Therefore, if a high percentage of questions in a topic do not have accepted an answer, we can understand that the topic is difficult.
Additionally, we analyzed the trends related to Ruby questions on SO over the years to understand how they have evolved.

\textbf{Step 7: Conducting the Survey.}
To see if we can support the results of our experimental studies, we implemented a $10$-minute survey using Qualtrics\footnote{https://www.qualtrics.com/}. The survey targeted Ruby developers and aimed to identify which aspects of the Ruby programming language Ruby developers found most challenging. All questions were mandatory, ensuring that participants completed each section before proceeding.

This study was approved by the Behavioral Research Ethics Board of the University of British Columbia.

The survey begins with three demographic questions designed to capture essential information about the participants. These questions help us understand the diversity of respondents by collecting data on their location, education level, and experience with Ruby programming.

The survey aimed to validate findings from our SO data analysis by capturing developers' subjective experiences with Ruby-related challenges. Participants were presented with $35$ topics, identified through our empirical study, and asked to rate each on a 5-point Likert scale for difficulty, with $5$ being the most challenging. They also estimated the time typically spent resolving issues in the topics they found challenging, using a Likert scale with options: less than $15$ minutes, between $15$ to $30$ minutes, between $30$ to $60$ minutes, between $1$ to $2$ hours, and more than $2$ hours. Participants reflected on their general experiences, rather than solving specific problems in real-time.

To validate our survey design, we conducted a pilot study with seven graduate students at the University of British Columbia who had experience with Python but not Ruby. Although they were not the target audience, the goal was to assess the clarity, structure, and usability of the survey to ensure it was well-designed for Ruby developers. The pilot study also helped identify potential issues, refine question wording, and estimate the completion time.

After that, we distributed the full survey via Prolific\footnote{https://www.prolific.co}, a platform previously used in similar research studies \cite{reid2022software, russo2022recruiting}. To ensure data quality, participants were pre-screened for proficiency in Ruby and prior programming experience. Since the survey took approximately $10$ minutes to complete, an attention-check question was included, requiring a correct response for the participant’s submission to be considered valid.
The survey was published on Prolific in August 2024, yielding a total of $211$ responses, of which $154$ were approved. Approved participants were automatically paid at a rate of \pounds 6 per hour following Prolific’s guidelines.
To analyze the survey responses, as \cite{liang2024large}, we report the percentage of participants who selected each option for every topic.

\section{Results}

In this section, we provide the results for the RQs.

\subsection{RQ1: Topics and Taxonomy}

Table \ref{table:1} shows the $35$ topics that we generated with BERTopic, $18$ middle categories that were generated with GPT-4o and manual analysis, and the six main categories of Ruby developer challenges. Most of the Ruby developer questions are about \textit{Web Application} Development ($27.55\%$ of all the questions). 
Web Application Development category is the process of creating dynamic and interactive applications that run on web browsers. This category covers diverse aspects of development including user interface and experience customization, frontend integration, asynchronous interactions with technologies like jQuery and AJAX, and backend concerns.

The next most discussed category is \textit{Application Quality and Security} ($23.28\%$). This category focuses on ensuring the reliability, and safety of web applications, addressing a range of concerns including handling method and parameter errors, validation and error-handling practices, mass assignment and parameter protection to safeguard against security vulnerabilities. Debugging errors and writing effective tests using tools like RSpec are also key areas of focus.

The third place is for questions related to \textit{Data Management and Processing} ($19.71\%$). This category encompasses a spectrum of tasks related to handling and organizing application data, such as optimizing data queries and manipulation within Rails, database management and schema design, ActiveRecord associations, JSON and data serialization problems, and questions on file handling and integrating with external services.

\textit{Development Environment and Infrastructure} is the fourth discussed category, encompassing $15.62\%$ of the discussions in our dataset. This category involves managing and configuring the various elements that support application development and deployment, such as job scheduling and background processing, deployment and server configuration issues, system integration and troubleshooting Heroku deployment issues, automation with tools like Chef, and managing Ruby environments and dependencies.

The fifth category in our derived taxonomy is \textit{Core Ruby Concepts} ($11.22\%$). This category includes foundational aspects of the Ruby programming language that are essential for Ruby development, including inquiries related to advanced Ruby methods and metaprogramming techniques, and core operations on Ruby arrays and hashes and regular expressions.

Only $2.59\%$ of the questions were related to \textit{Software Architecture and Performance}. The questions in this category are related to enhancing the efficiency and design of Ruby applications, designing pattern queries, and improving modularity across multiple Rails applications.

\subsection{RQ2: Analyzing the Topics}

\textbf{Popularity:} 
Table \ref{table:1} shows the popularity metrics for $35$ topics.
\textit{Advanced Ruby Methods and Metaprogramming} has the highest average scores, indicating strong developer interest, especially in areas like managing thread safety and implementing custom setters.
\textit{Ruby Array and Hash Operations}, which has the highest average views, covers essential tasks like iterating and manipulating arrays. Since array and hash manipulations are central to many Ruby applications, this topic likely appeals to developers at all skill levels, from beginners to experts.
\textit{Ruby Environment and Dependency Management} closely follows in popularity, addressing key challenges like accessing Rails environments and resolving gem version conflicts.
\textit{Time Zone Management and Date Operations in Rails} is also popular, especially for its coverage of time zone conversions which is a frequent need in many applications.

\textbf{Difficulty:}
Table \ref{table:1} shows three metrics for difficulty.
\textit{Ruby Gem Installation and Configuration Issues} has the highest percentage of questions without an accepted answer, suggesting that developers encounter challenges when setting up and managing gems in Rails applications. Questions about \textit{Search Engines Integration} have longer response times, reflecting the complex work involved in configuring advanced search functionalities and managing hierarchical relationships in Elasticsearch. These topics seem to pose unique difficulties, likely due to their technical complexity and critical role in application functionality.

\definecolor{cat1}{HTML}{fbb4ae}
\definecolor{cat2}{HTML}{b3cde3}
\definecolor{cat3}{HTML}{FDE8B2}
\definecolor{cat4}{HTML}{ccebc5}
\definecolor{cat5}{HTML}{decbe4}
\definecolor{cat6}{HTML}{f1e2cc}

\definecolor{cat11}{HTML}{f77f85} 
\definecolor{cat12}{HTML}{6baed6} 
\definecolor{cat13}{HTML}{f9d670} 
\definecolor{cat14}{HTML}{82b88b} 
\definecolor{cat15}{HTML}{a68fba} 
\definecolor{cat16}{HTML}{b8a17e} 

\begin{table*}[ht]
\centering
\resizebox{\textwidth}{!}{
\begin{tabular}{lccccccc}
\toprule
\multirow{2}{*}{\textbf{Topic}} & \multicolumn{2}{c}{\textbf{Popularity Metrics}} & \multicolumn{3}{c}{\textbf{Difficulty Metrics}} & \multirow{2}{*}{\textbf{\# of Questions}} & \multirow{2}{*}{\textbf{Middle Category}} \\ 
\cmidrule(lr){2-3} \cmidrule(lr){4-6}
& \textbf{View} & \textbf{Score} & \textbf{w/o acc. (\%)} & \textbf{w/o ans. (\%)} & \textbf{Mins to acc.} & & \\
\midrule

\rowcolor{cat1} 
ActiveRecord Associations in Rails & 1585.17 & 1.74 & 37.37 & 7.24 & 37 & 28809 &  \\
\rowcolor{cat1} 
Data Query and Manipulation in Rails & 1986.21 & 2.28 & 36.42 & 6.59 & 34 & 18961 &  \\ 
\rowcolor{cat1} 
Database Management and Schema Design & 2465.54 & 2.97 & 44.24 & 11.25 & 68 & 22859 &  \multirow{-3}{*}{Database and ActiveRecord}\\
\arrayrulecolor{cat1} \specialrule{6pt}{0pt}{-6pt} \arrayrulecolor{black}
\cmidrule{8-8}

\rowcolor{cat1} 
JSON and Data Serialization & 2144.73 & 2.36 & 35.39 & 6.92 & 48 & 10602 &  \\
\rowcolor{cat1} 
Spreadsheet and CSV Management in Ruby & 2235.47 & 2.04 & 42.27 & 9.25 & 63 & 5592 &  \\ 
\rowcolor{cat1} 
File Handling and External Integrations & 1762.58 & 2.05 & 51.98 & 15.14 & 221 & 10332 &  \multirow{-3}{*}{Data Handling and Serialization}\\ 
\midrule

\rowcolor{cat2} Advanced Ruby Methods and Metaprogramming & 3223.39 & \textbf{5.39} & 25.87 & 2.87 & 17 & 20921 & Advanced Ruby Concepts \\
\arrayrulecolor{cat2} \specialrule{6pt}{0pt}{-6pt} \arrayrulecolor{black}
\cmidrule{8-8}
\rowcolor{cat2} Algorithm Design in Ruby & 1877.90 & 2.41 & 28.20 & 3.14 & 16 & 11785 & Algorithm Design \\ 
\arrayrulecolor{cat2} \specialrule{6pt}{0pt}{-6pt} \arrayrulecolor{black}
\cmidrule{8-8}
\rowcolor{cat2} Regular Expressions in Ruby & 2096.35 & 2.29 & 22.80 & 1.92 & 14 & 5092 &  \\
\rowcolor{cat2} Ruby Array and Hash Operations & \textbf{4301.20} & 4.45 & 25.00 & 2.02 & 16 & 17510 & \multirow{-2}{*}{Core Ruby Operations}\\
\midrule

\rowcolor{cat3} API Management & 1627.43 & 1.96 & 51.59 & 14.60 & 188 & 17717 &  \\
\rowcolor{cat3} Asset Management and Integration Issues & 1810.65 & 2.52 & 51.78 & 15.79 & 199.5 & 9265 & \\
\rowcolor{cat3} Email Delivery with Rails ActionMailer & 1767.06 & 2.11 & 48.83 & 12.75 & 105 & 5309 & \\
\rowcolor{cat3} Ruby Payment and Financial Integration & 1139.70 & 1.43 & 54.22 & 14.34 & 334 & 2358 & \\
\rowcolor{cat3} Search Engines Integration in Rails & 994.21 & 1.33 & 48.04 & 13.90 & \textbf{428} & 3488 & \multirow{-5}{*}{External Services and API Integrations} \\
\arrayrulecolor{cat3} \specialrule{6pt}{0pt}{-6pt} \arrayrulecolor{black}
\cmidrule{8-8}
\rowcolor{cat3} Frontend Integration and User Interaction in Rails Applications & 1753.48 & 1.42 & 44.42 & 10.61 & 49 & 15565 &  \\
\rowcolor{cat3} jQuery and AJAX in Rails & 1351.68 & 1.25 & 46.18 & 12.63 & 68 & 23446 & \\
\rowcolor{cat3} Ruby on Rails Web Interface and UX Customization & 1936.53 & 1.93 & 38.62 & 7.69 & 37 & 29237 & \multirow{-3}{*}{Frontend Development in Rails} \\
\arrayrulecolor{cat3} \specialrule{6pt}{0pt}{-6pt} \arrayrulecolor{black}
\cmidrule{8-8}
\rowcolor{cat3} Rails Routing and URLs & 1681.20 & 2.01 & 35.40 & 6.03 & 29 & 11589 & Routing and URLs \\
\arrayrulecolor{cat3} \specialrule{6pt}{0pt}{-6pt} \arrayrulecolor{black}
\cmidrule{8-8}
\rowcolor{cat3} Time Zone Management and Date Operations in Rails & 3667.03 & 3.84 & 33.91 & 5.40 & 27 & 4406 & Time and Date Management \\
\arrayrulecolor{cat3} \specialrule{6pt}{0pt}{-6pt} \arrayrulecolor{black}
\cmidrule{8-8}
\rowcolor{cat3} User Authentication and Role Management & 1742.13 & 2.36 & 48.66 & 12.78 & 114 & 13463 & User Authentication \\
\midrule

\rowcolor{cat4} Automated Testing and Integration Testing in Ruby on Rails & 1982.34 & 2.68 & 44.15 & 10.30 & 130 & 29447 & Testing and Quality Assurance \\
\arrayrulecolor{cat4} \specialrule{6pt}{0pt}{-6pt} \arrayrulecolor{black}
\cmidrule{8-8}
\rowcolor{cat4} Rails Mass Assignment \& Parameter Protection Issues & 2109.43 & 2.75 & 37.06 & 7.65 & 40 & \textbf{31271} &  \\
\rowcolor{cat4} Rails Method and Parameter Errors & 1504.43 & 1.42 & 42.49 & 10.85 & 38 & 27971 & \\
\rowcolor{cat4} Ruby Debugging and Error Handling & 1710.78 & 1.81 & 48.17 & 14.22 & 51 & 15628 & \\
\rowcolor{cat4} Validation and Error Handling in Ruby on Rails & 1875.07 & 2.24 & 36.90 & 7.29 & 35 & 10468 & \multirow{-4}{*}{Error Handling and Validation} \\
\midrule

\rowcolor{cat5} Automation with Chef in DevOps & 1708.75 & 1.30 & 46.25 & 8.22 & 251 & 1412 &  \\
\rowcolor{cat5} Heroku Deployment and Configuration & 1645.78 & 2.16 & 50.22 & 13.53 & 140 & 8081 & \\
\rowcolor{cat5} Rails Deployment and Server Configuration & 1777.89 & 2.17 & 52.12 & 14.91 & 252.5 & 7611 & \multirow{-3}{*}{Deployment and Infrastructure Management} \\
\arrayrulecolor{cat5} \specialrule{6pt}{0pt}{-6pt} \arrayrulecolor{black}
\cmidrule{8-8}
\rowcolor{cat5} Job Scheduling and Background Processes & 1908.61 & 2.51 & 49.78 & 13.31 & 162 & 6820 & Background Processing \\
\arrayrulecolor{cat5} \specialrule{6pt}{0pt}{-6pt} \arrayrulecolor{black}
\cmidrule{8-8}
\rowcolor{cat5} Ruby Environment and Dependency Management & 3499.33 & 3.92 & 47.33 & 11.10 & 94 & 23762 &  \\
\rowcolor{cat5} Ruby Gem Installation and Configuration Issues & 2691.02 & 2.82 & \textbf{54.88} & 15.44 & 183 & 14488 & \multirow{-2}{*}{Environment and Dependency Management} \\
\arrayrulecolor{cat5} \specialrule{6pt}{0pt}{-6pt} \arrayrulecolor{black}
\cmidrule{8-8}
\rowcolor{cat5} System Integration and External Libraries in Ruby & 2361.13 & 3.16 & 42.64 & 10.82 & 79 & 16245 & System Integration \\
\midrule

\rowcolor{cat6} Performance Optimization in Rails & 1270.92 & 2.48 & 49.76 & \textbf{16.03} & 117 & 4660 & Performance Optimization \\
\arrayrulecolor{cat6} \specialrule{6pt}{0pt}{-6pt} \arrayrulecolor{black}
\cmidrule{8-8}
\rowcolor{cat6} Rails Design Patterns & 1026.79 & 2.09 & 37.11 & 7.32 & 48 & 8151 & Software Design Patterns \\
\midrule
\textbf{Average for all Ruby questions} & 2092.46 & 2.49 & 41.60 & 9.68 & 48 & \textbf{Total: 494321} & - \\
\bottomrule
\end{tabular}
}
\caption{The three-level hierarchy and popularity and difficulty metrics for $35$ Ruby topics. (w/o acc.): percentage of questions without accepted answer, (w/o ans.): percentage of questions without an answer, (Mins to acc.): median time for a question to receive an accepted answer. The colors represent the six top-level categories: \textcolor{cat11}{Data Management and Processing}, \textcolor{cat12}{Core Ruby Concepts}, \textcolor{cat13}{Web Application Development}, \textcolor{cat14}{Application Quality and Security}, \textcolor{cat15}{Development Environment and Infrastructure}, \textcolor{cat16}{Software Architecture and Performance}.} 
\label{table:1}
\end{table*}

\begin{figure}[htbp]
\includegraphics[width=\columnwidth]{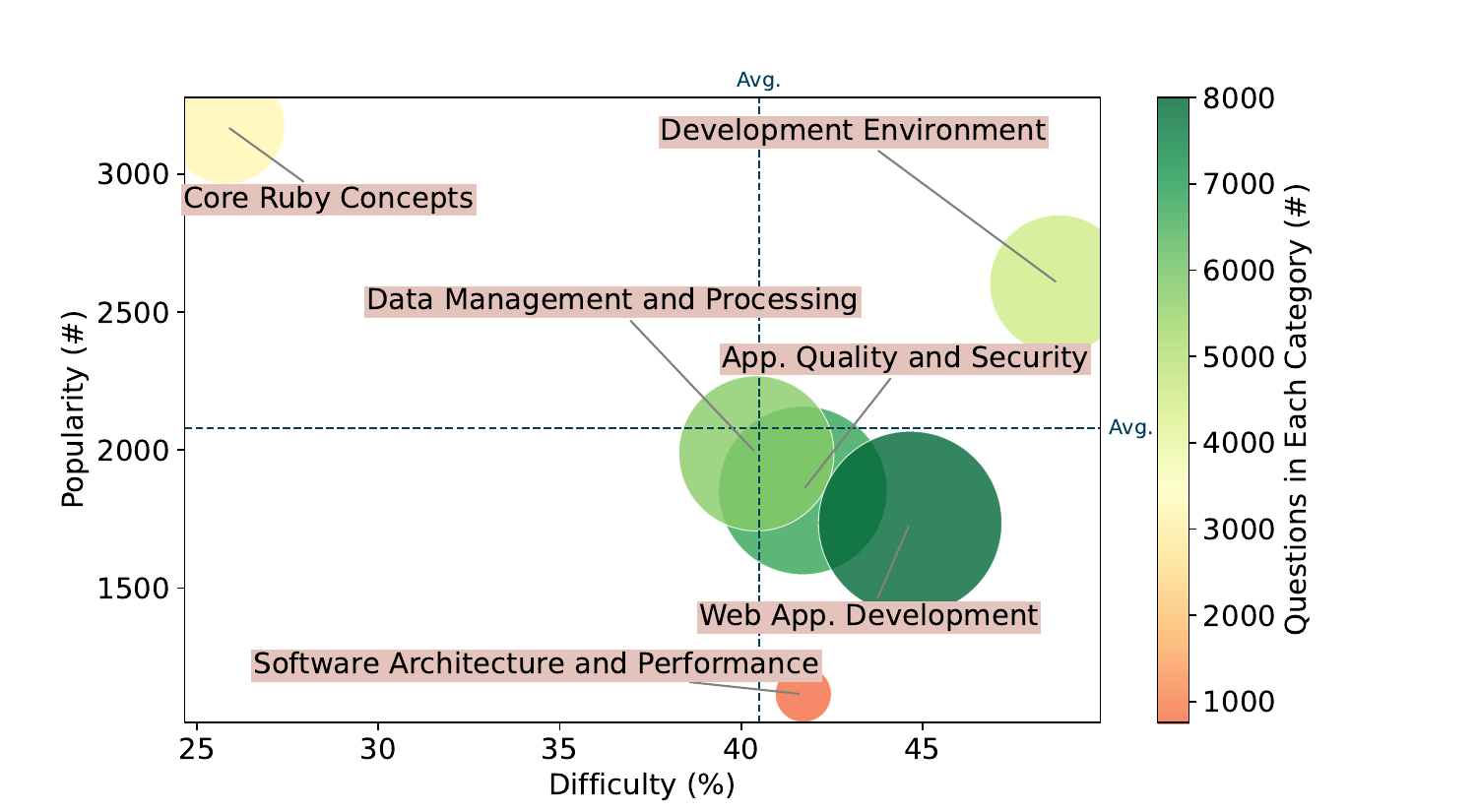}
\centering
\caption{Popularity vs Difficulty.}
\label{fig:popularity_vs_difficulty}
\end{figure}

Figure \ref{fig:popularity_vs_difficulty} illustrates the trade-off between popularity (average number of views) and difficulty (percentage of questions without accepted answers) for each of the six main categories based on SO data. 
The size and color of the circles indicate the number of questions within each category. \textit{Core Ruby Concepts} are highly popular but not considered difficult. This is likely because these concepts are essential for any Ruby developer and serve as foundational knowledge for those starting with Ruby programming, resulting in relatively straightforward questions.
The \textit{Development Environment and Infrastructure} category emerges as the most difficult while also exhibiting above-average popularity. \textit{Software Architecture and Performance} is the least popular category, yet it has above-average difficulty. The other three categories are characterized by both moderate popularity and difficulty. This suggests that while questions in these areas are frequently viewed, they present challenges for developers.

To understand whether the more difficult topics are also more popular,
we conducted a Kendall rank correlation test at a $95$\% confidence level \cite{hussain2019pymannkendall}, adopted from prior research \cite{ahmed2018concurrency}. 
The \textbf{null hypothesis} ($H_0$) is that there is no significant relationship between average popularity and average difficulty, and the \textbf{alternative hypothesis} ($H_1$) is that there is a significant relationship between average popularity and average difficulty.
The Kendall correlation coefficient was found to be $-0.33$, indicating a moderate negative correlation between popularity and difficulty. However, the $p$-value of $0.47$ suggests that this correlation is not statistically significant. Thus, we cannot confidently assert that a meaningful relationship exists between average popularity and average difficulty.

\begin{figure}[htbp]
\centering
\begin{subfigure}[t]{0.5\columnwidth}
    \centering
    \includegraphics[width=\linewidth]{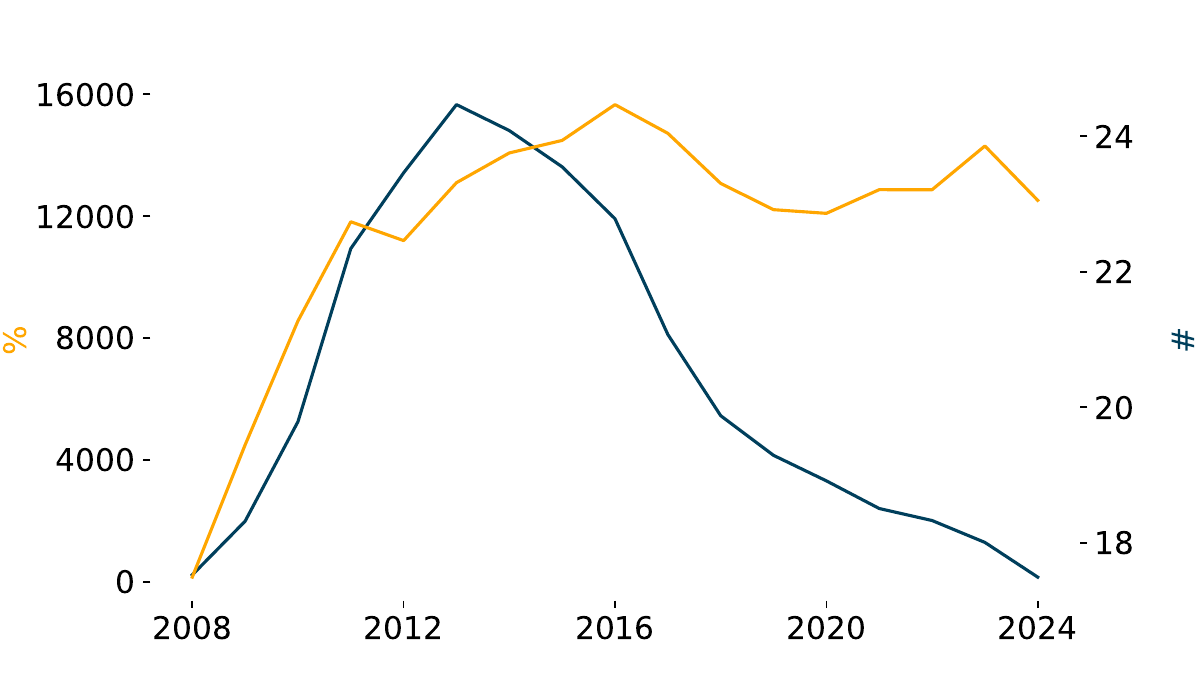}
    \caption{Application Quality.}
    \label{fig:3a}
\end{subfigure}%
\hfill
\begin{subfigure}[t]{0.5\columnwidth}
    \centering
    \includegraphics[width=\textwidth]{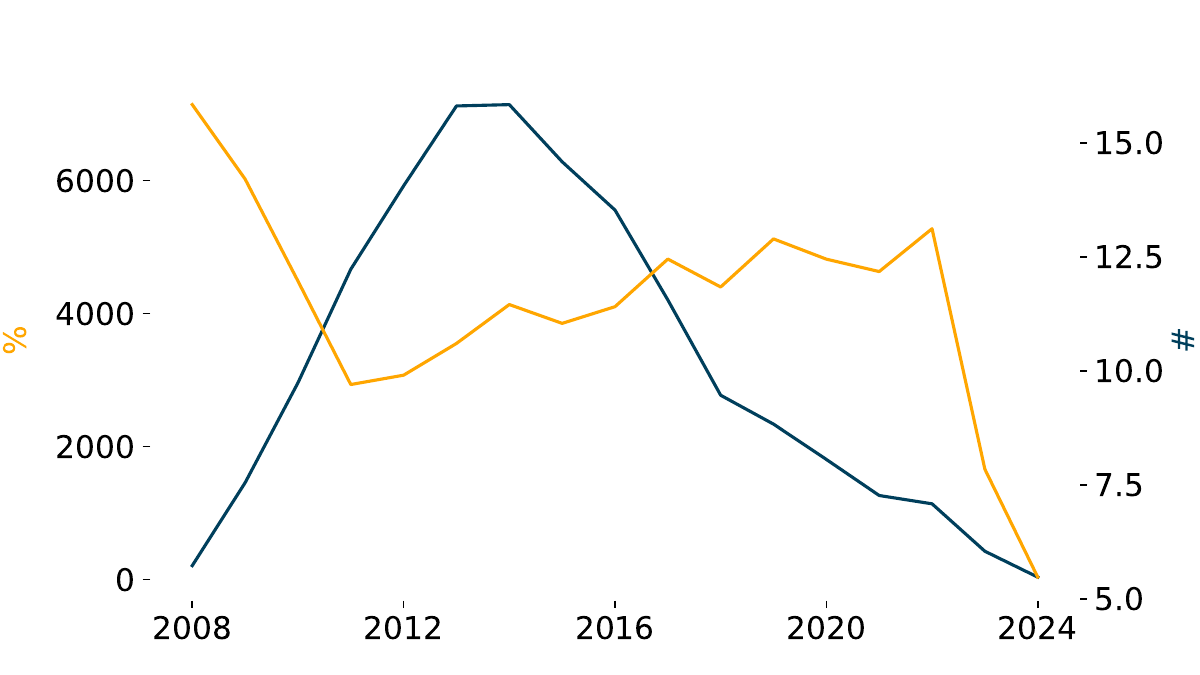}
    \caption{Core Ruby Concepts.}
    \label{fig:3b}
\end{subfigure}
\vspace{1em} 
\begin{subfigure}[t]{0.5\columnwidth}
    \centering
    \includegraphics[width=\linewidth]{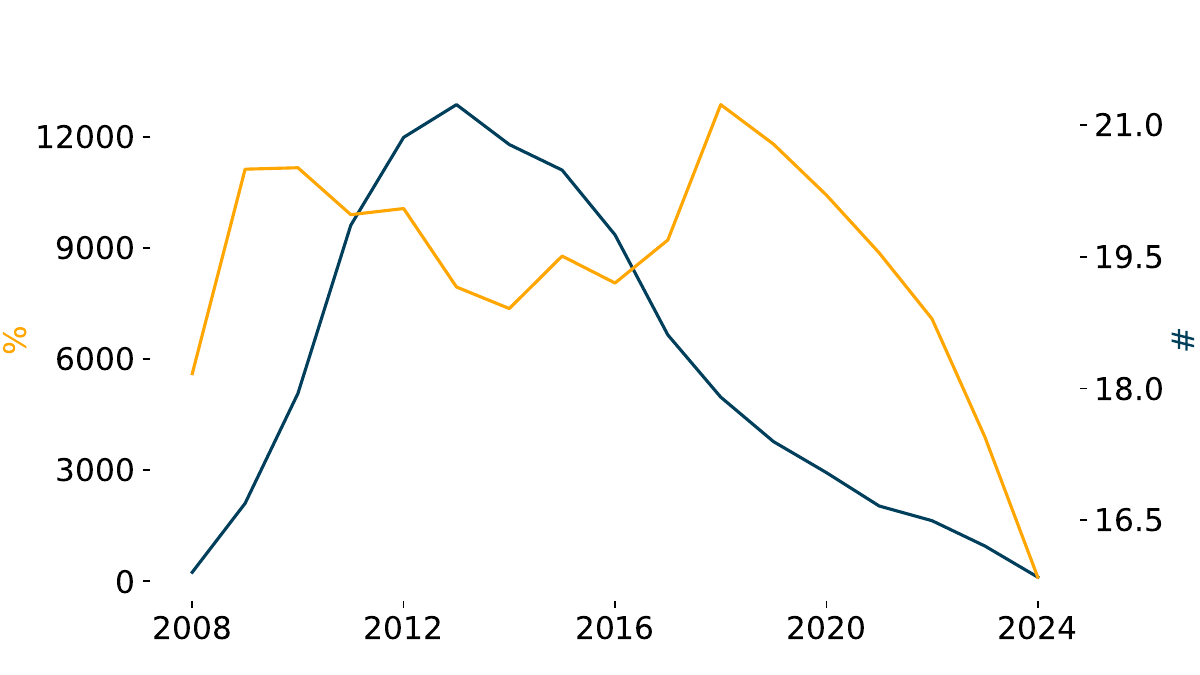}
    \caption{Data Management.}
    \label{fig:3c}
\end{subfigure}%
\hfill
\begin{subfigure}[t]{0.5\columnwidth}
    \centering
    \includegraphics[width=\linewidth]{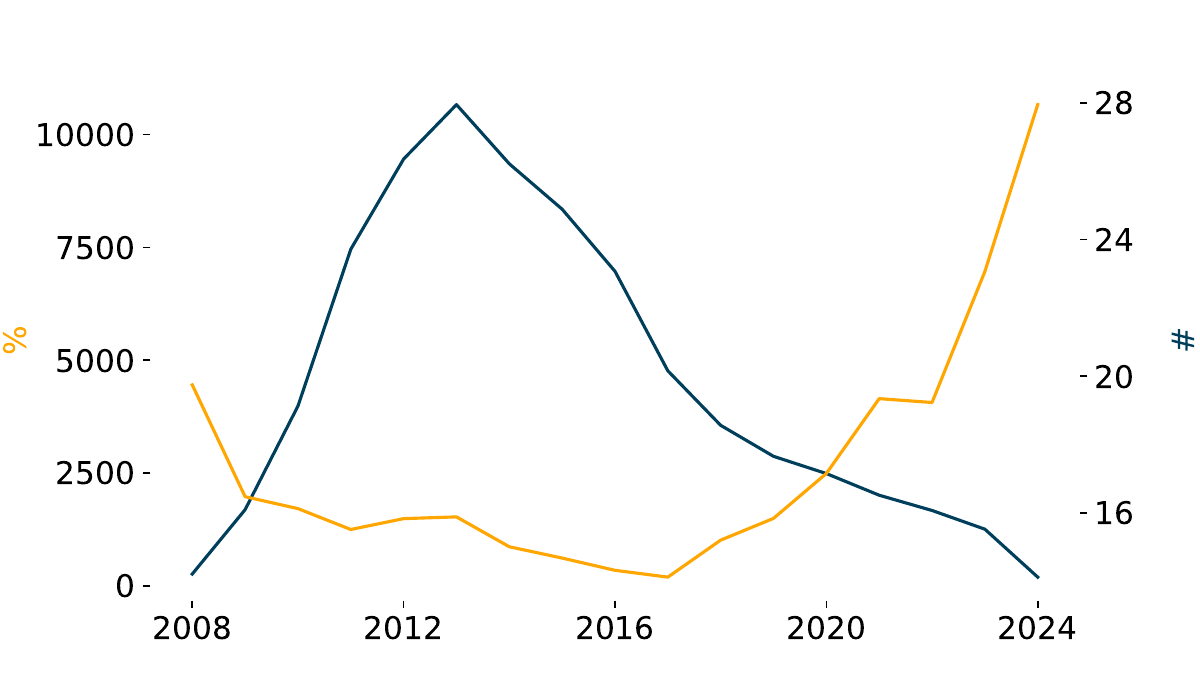}
    \caption{Development Environment.}
    \label{fig:3d}
\end{subfigure}
\vspace{1em} 
\begin{subfigure}[t]{0.5\columnwidth}
    \centering
    \includegraphics[width=\linewidth]{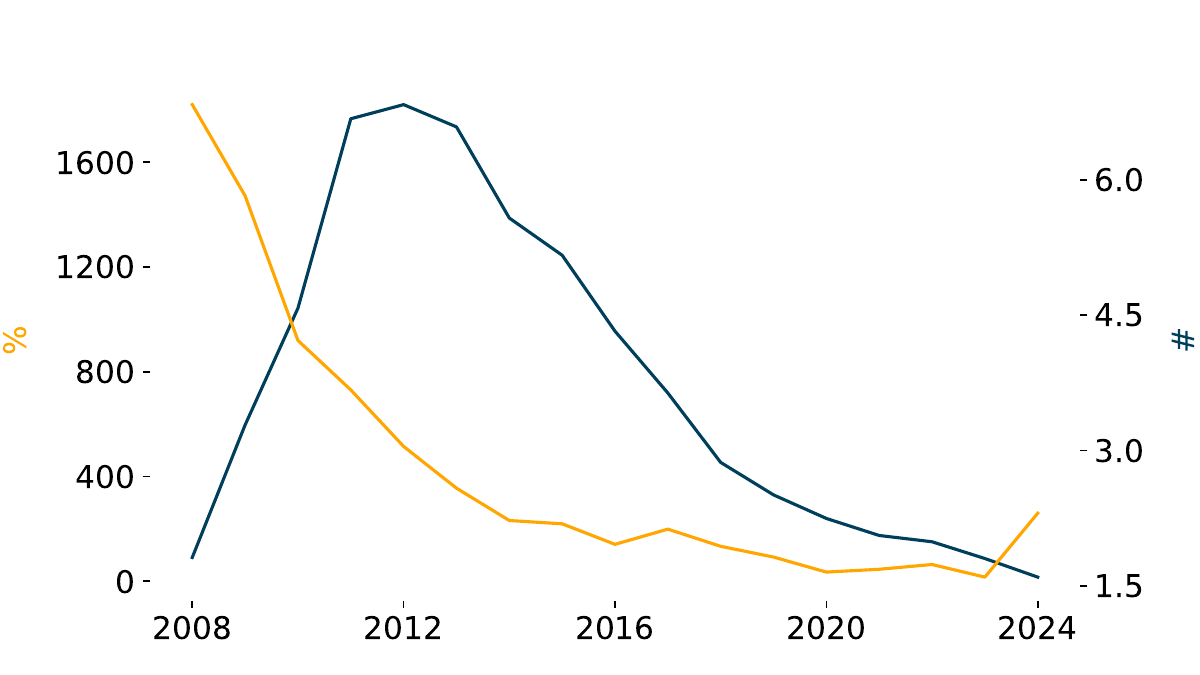}
    \caption{Software Architecture.}
    \label{fig:3e}
\end{subfigure}%
\hfill
\begin{subfigure}[t]{0.5\columnwidth}
    \centering
    \includegraphics[width=\linewidth]{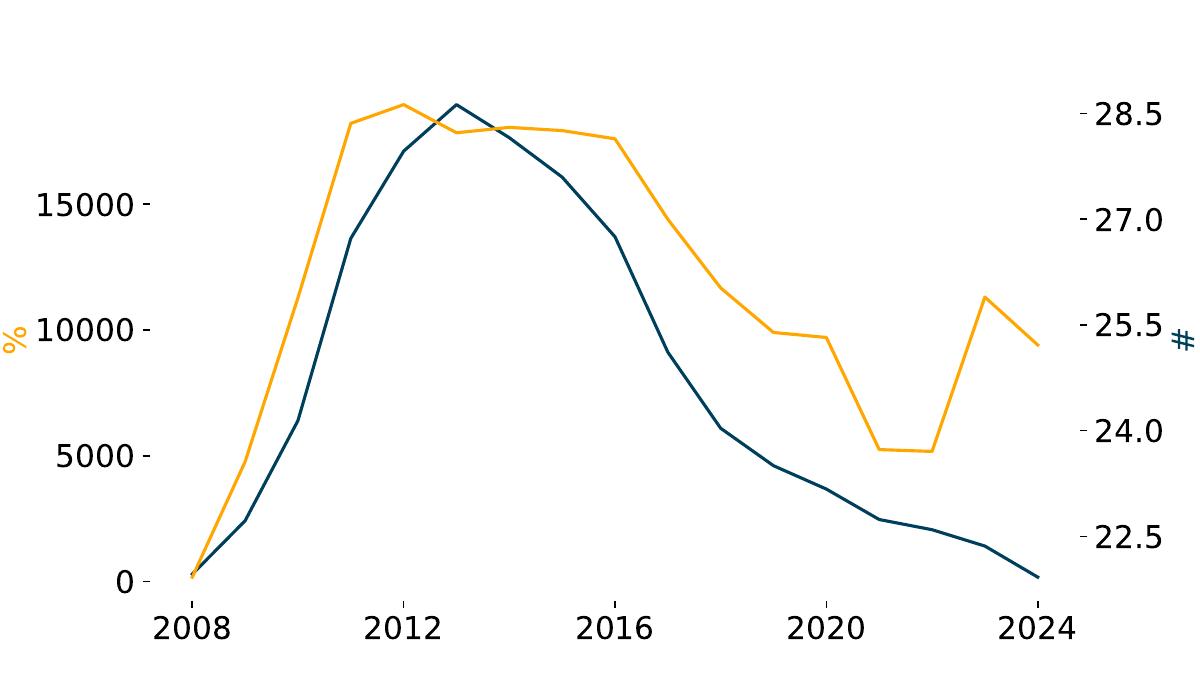}
    \caption{Web Application.}
    \label{fig:3f}
\end{subfigure}
\caption{Trends of six main categories over time. \textcolor[HTML]{003f5c}{Blue}: total number of category-specific questions over the years; \textcolor[HTML]{ffa600}{Yellow}: percentage of category-specific questions relative to all Ruby-related questions. }
\label{fig:category_trends}
\end{figure}

\textbf{Trends:} To analyze the trends in Ruby-related questions across six main categories, we examined two key metrics: the total number of questions over time and the percentage of category-specific questions relative to all Ruby-related questions. Figure \ref{fig:category_trends} illustrates these trends for each category. From the graph, it is evident that the number of questions for all categories grew steadily from $2008$ until approximately $2013$. However, after this period, a decline is observed across all categories, continuing until the present day.

The percentage of questions in the \textit{Software Architecture and Performance} category was notably high initially but has steadily decreased since. On the other hand, the \textit{Application Quality and Security} category percentage of questions saw a consistent rise in its share of questions, stabilizing after a certain point. The \textit{Core Ruby Concepts} category percentage of questions initially declined, then experienced a period of increase, but began to decrease again around $2023$. The \textit{Data Management and Processing category} experienced growth early on, but after $2019$, its percentage started to decline. Finally, the \textit{Web Application Development} category percentage of questions initially saw an increase, followed by a stable period, then a decline. However, by $2022$, this category began to rise again.

To analyze the trends, we employed the Mann-Kendall test alongside the percentage of questions in each category relative to all Ruby-related questions \cite{hussain2019pymannkendall}. The Mann-Kendall test is a non-parametric test to detect trends in time series and is used in previous studies \cite{hussain2019pymannkendall, haque2020challenges}. 
We applied a standard $95\%$ confidence level to determine whether each topic is exhibiting a statistically significant increase or decrease over time. The results are summarized in Table \ref{tab:mankendall_results}.
\textit{Application Quality and Security} and \textit{Development Environment and Infrastructure} demonstrated statistically significant increasing trends. This suggests a growing interest and engagement in these areas over time, as reflected by the increasing number of questions being asked.
In contrast, both the \textit{Software Architecture and Performance} and \textit{Web Application Development} categories exhibited significant decreasing trends, indicating a notable decline in engagement in these domains.
The test results show no significant trend for the other two topics.
 
This stability implies that the percentage of questions related to these categories remained consistent over the observed period, reflecting a steady level of engagement without substantial fluctuations for \textit{Core Ruby Concepts} and \textit{Data Management and Processing}.

\begin{table}[ht]
    \centering
    \caption{Mann-Kendall test results for main categories trends. }
    \begin{tabular}{@{}llll@{}}
        \toprule
        \textbf{Category} & \textbf{Z Value} & \textbf{P Value} & \textbf{Trend} \\ \midrule
        Application Quality & 5.86 & $4.63 \times 10^{-9}$ & Increasing \\
        Core Ruby Concepts & 0.65 & 0.51 & No Trend \\
        Data Management & -1.52 & 0.13 & No Trend \\
        Development Environment & 4.11 & $3.93 \times 10^{-5}$ & Increasing \\
        Software Architecture & -13.76 & 0.0 & Decreasing \\
        Web Application & -4.67 & $3.05 \times 10^{-6}$ & Decreasing \\ \bottomrule
    \end{tabular}
    \label{tab:mankendall_results}
\end{table}

\subsection{RQ3: Survey Results-Difficulty}

\begin{figure}[htbp]
\includegraphics[width=\columnwidth]{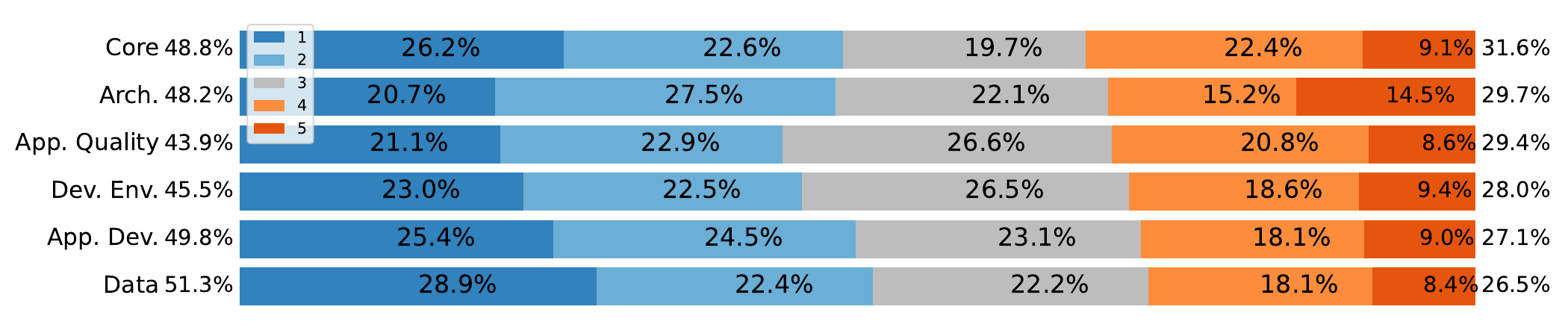}
\centering
\caption{Survey results of the difficulty levels divided by each category: (1)not difficult, (2)slightly difficult, (3)moderate, (4)difficult, (5)very difficult.}
\label{fig:survey_part1}
\end{figure}

Participants were categorized based on their Ruby development experience: those with less than two years were considered novice developers, while those with more than six years of experience were classified as experienced developers. Among the respondents, we identified $68$ novice developers, with $18$ of them having less than a year of experience, and $36$ experienced developers, which included $12$ individuals with over $10$ years of experience in Ruby development.
Figure \ref{fig:survey_part1} presents the normalized results of the survey question for the difficulty level of each main category. 

The percentages on the right-hand side illustrate the proportion of respondents who rated each category as difficult (4) or very difficult (5), while those on the left-hand side represent the percentage of participants who found the category not difficult at all (1) or slightly difficult (2).
The survey results indicate that \textit{Core Ruby Concepts} is the most challenging category, with $31.6\%$ of participants rating it as difficult. In contrast, \textit{Data Management and Processing} was perceived as the easiest category, with $51.3\%$ of respondents finding it not difficult.
Among experienced developers, $42.13\%$ reported that they found \textit{Application Quality and Security} to be difficult, while $49.29\%$ indicated that they did not find \textit{Core Ruby Concepts} challenging.
For novice developers, $26.54\%$ expressed that they struggle with \textit{Software Architecture and Performance}, whereas $54.71\%$ stated that \textit{Data Management and Processing} does not present difficulties for them.
The perceived difficulty of \textit{Core Ruby Concepts} versus \textit{Data Management and Processing} likely stems from their nature. \textit{Core Ruby Concepts} involve advanced features like metaprogramming and complex methods, which are not essential for routine data management. In contrast, \textit{Data Management and Processing} is often simplified through frameworks and gems. While experienced developers found \textit{Core Ruby Concepts} less difficult than novices, both groups reported similar difficulty with \textit{Data Management and Processing}, suggesting that data management is more accessible due to available tools and abstractions.

To assess how well the survey findings aligned with Stack Overflow (SO) data, we first examined three key difficulty metrics from SO across $35$ topics. Since our survey captures subjective perceptions of difficulty using ordinal Likert scale responses (1–5), while SO provides continuous numerical data based on user interactions, a direct comparison would be inappropriate. After consulting with a professor specializing in statistics and machine learning, we determined that a regression model would allow us to meaningfully analyze the relationship between these variables. This approach enabled us to standardize both datasets to a common scale, facilitating direct comparisons and helping us determine whether higher perceived difficulty in the survey corresponded to higher difficulty indicators in SO data.
Before constructing the regression model, we conducted a correlation analysis to identify the most relevant SO metrics. Our analysis revealed a strong positive correlation ($0.97$) between the percentage of questions without an accepted answer and the percentage of unanswered questions. Additionally, moderate positive correlations were observed between the percentage of questions without an accepted answer and the median time to receive an accepted answer ($0.66$), as well as between the percentage of unanswered questions and the median time to receive an accepted answer ($0.70$). Given these relationships, we selected the percentage of questions without an accepted answer and the median response time as predictors in our regression model, ensuring that we captured distinct yet related aspects of difficulty.
To further validate our model, we performed a Variance Inflation Factor (VIF) analysis to check for multicollinearity, which occurs when independent variables are highly correlated, which can compromise regression estimates and reduce model reliability. Following established guidelines in statistical analysis, we used the common threshold of $5$ for VIF scores, where values exceeding this threshold suggest problematic multicollinearity \cite{o2007caution}.
For each of the $35$ topics, we computed a weighted mean difficulty score from the survey responses to align with SO metrics. Finally, we standardized both datasets to a mean of $zero$ and a standard deviation of $one$, ensuring consistency in scale and facilitating meaningful comparisons between perceived and objective measures of difficulty.

We formulated our hypotheses as follows: the \textbf{null hypothesis} ($H_0$) states that there is no significant relationship between the percentage of unanswered questions on SO and the perceived difficulty of topics, while the \textbf{alternative hypothesis} ($H_1$) suggests a significant relationship exists.
To test these hypotheses, we employed a multiple regression model to examine the relationship between the survey-derived perceived difficulty measure and two SO-derived metrics: the percentage of questions without an accepted answer and the median time for questions to receive an accepted answer. The analysis produced a slope of $0.016$ for the percentage of questions without an accepted answer and $0.14$ for median response time, with an intercept near $zero$. The model’s $R^2$ value of $0.022$ indicates that these predictor variables account for only $2.2\%$ of the variation in the perceived difficulty measure.

Given the regression results, we fail to reject the null hypothesis.
From this model, it appears that developers' perceptions do not align with SO data. The low amount of variation explained by the model ($R^2=0.022$) suggests that the perception metrics on difficulty and difficulty levels on SO are not linearly related. 
Thus, the survey data, which reflects subjective ratings, and SO data, based on objective platform metrics, are not strongly correlated. 
This discrepancy implies that the difficulty that developers experience may not match the difficulty indicated by SO metrics, meaning we cannot reliably predict one type of difficulty measure from the other.
In conclusion, the results suggest that subjective perceptions and objective metrics provide distinct perspectives on difficulty, and each may capture different aspects of the challenges developers face.

\subsection{RQ4: Survey Results-Time}

Figure \ref{fig:survey_part2} demonstrates the normalized results for the time it takes for survey participants to resolve issues. 
 
On the right side of the chart, we can see the percentage of respondents who reported spending more than $1$ hour to resolve their problems, while the left-hand side reflects the percentages of those who managed to resolve their issues in less than $30$ minutes.
The findings indicate that within the \textit{Software Architecture and Performance} category, $29.5\%$ of participants required more than one hour to address their issues, which is the highest percentage of any category for those taking more than $2$ hours at $13.4\%$. In contrast, the \textit{Application Quality and Security} category revealed that $44.9\%$ of participants resolved their issues in under $30$ minutes. Furthermore, the \textit{Web Application Development} category exhibited the highest percentage of participants, at $16.7\%$, who were able to resolve their problems in less than $15$ minutes.

\definecolor{a1}{HTML}{1F4E7C}
\definecolor{a2}{HTML}{4F8FB3}
\definecolor{a3}{HTML}{585858}
\definecolor{a4}{HTML}{F77A3C }
\definecolor{a5}{HTML}{B43A08}

\begin{figure}[htbp]
\includegraphics[width=\columnwidth]{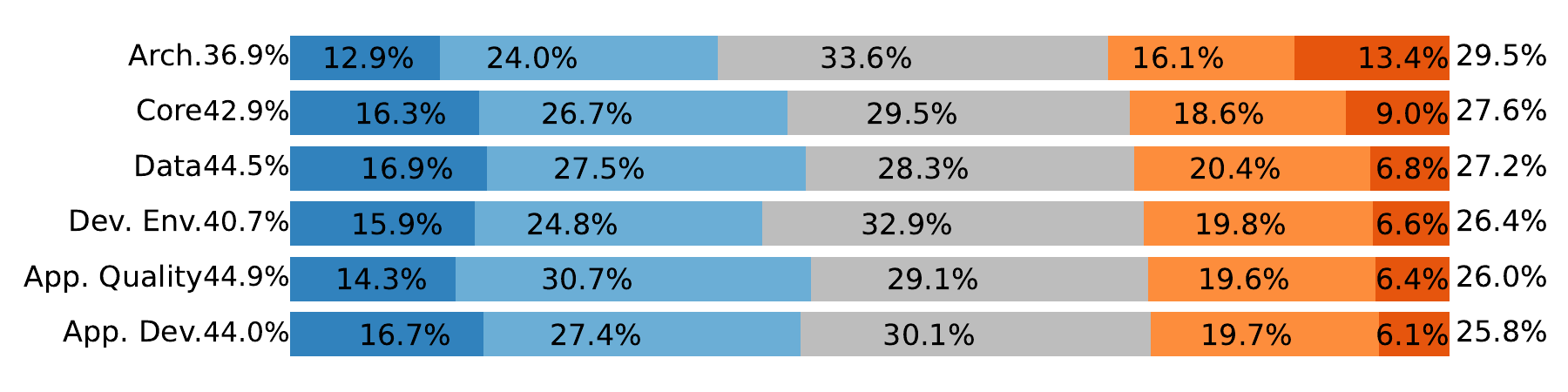}
\centering
\caption{Survey results of the required time to resolve issues divided by each category: \textcolor{a1}{less than $15$ min}, \textcolor{a2}{between $15$ to $30$ min}, \textcolor{a3}{between $30$ to $60$ min}, \textcolor{a4}{between $1$ and $2$ hours}, \textcolor{a5}{more than $2$ hours}. }
\label{fig:survey_part2}
\end{figure}

\begin{figure}[htbp]
\includegraphics[width=\columnwidth]{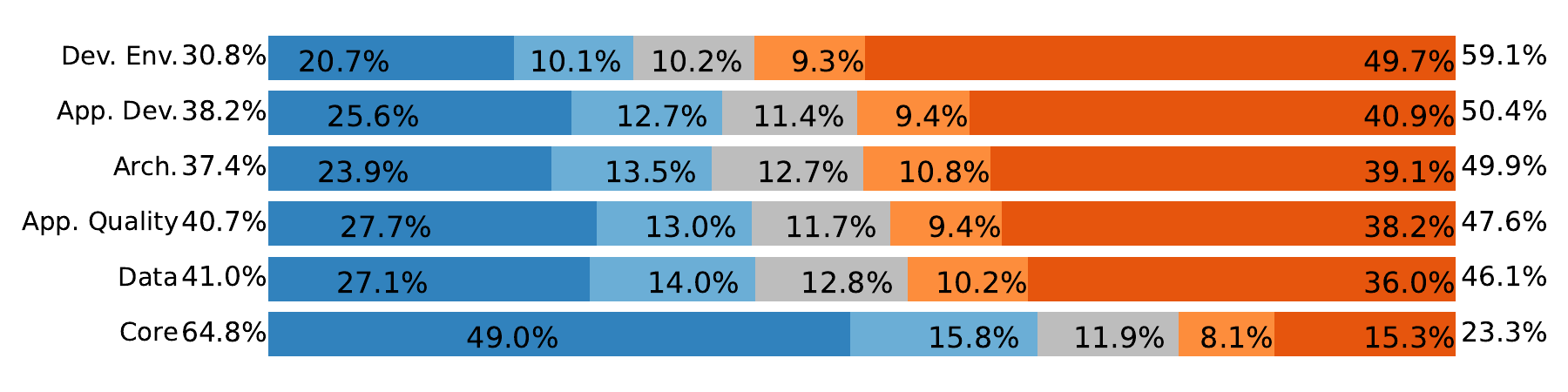}
\centering
\caption{ SO results of the time to receive an accepted answer by each category: \textcolor{a1}{less than $15$ min}, \textcolor{a2}{between $15$ to $30$ min}, \textcolor{a3}{between $30$ to $60$ min}, \textcolor{a4}{between $1$ and $2$ hours}, \textcolor{a5}{more than $2$ hours}.}
\label{fig:time_to_receive_accepted_answer_SO}
\end{figure}

To assess the alignment of the survey results with the experimental study data, a chart representing the time taken to receive an accepted answer based on SO data was created and is shown in Figure \ref{fig:time_to_receive_accepted_answer_SO}.
The time for questions to receive an accepted answer in the SO data was mapped to the five categorical options used in the survey. This mapping allows a direct comparison of the two data sets, aligning the survey responses with the actual SO data.

Although the comparison of the charts reveals that the survey results have a different distribution from the SO data, a $chi$-squared goodness of fit test was also conducted to statistically evaluate whether the observed differences between the distributions were significant.
This test is used to determine if there is a significant difference between the expected and observed frequencies in categorical data. In this case, the test compares the distribution of the time it took for survey participants to resolve their issues with the time it took for questions to receive an accepted answer on SO, categorized into five intervals.
The \textbf{null hypothesis ($H_0$)} for the $chi$-squared test states that there is no significant difference between the distribution of survey responses and the distribution observed in the SO data. Conversely, the \textbf{alternative hypothesis ($H_1$)} asserts that there is a significant difference between the two distributions, meaning that the survey responses do not align with the actual times on SO.
The test results for all categories indicate significant differences between the distributions of survey responses and the SO data. Specifically, for all six main categories, the $chi$-squared statistics were substantially high, with a $p$-value of $0$.

Consequently, we reject the null hypothesis, confirming that the survey data on time-to-resolution does not align with SO’s time-to-resolution metric. 

This discrepancy between RQ3 and RQ4 results highlights that developers perceive difficulty and time to resolution differently from what SO data indicates.
This could be due to the data from SO and the survey being inherently different and not comparable.
Another indication could be similar to the well-known ``stated vs. revealed preferences'' in behavioral sciences \cite{de2021stated}; suggesting a potential need for alternative or additional metrics to better capture the complexity and difficulty of issues from a developer’s perspective and SO. 
As our model shows, the variation in difficulty level is not well explained by time-to-resolution, meaning that time-to-resolution alone may not accurately reflect the difficulty that developers experience.
Given the low $R^2$ value in our model, this finding likely points to a need for improved or supplementary metrics on SO, as current metrics seem insufficient for capturing perceived difficulty levels.

\section{Discussions}

The findings of this study offer valuable insights for Ruby developers, tool creators, AI researchers, and the broader research community. The three-level taxonomy developed in RQ1 provides a structured framework for developers to assess their skills and identify areas for improvement. By highlighting the most difficult and popular topics, it serves as a road map for targeted learning and improved documentation.

In our trend analysis from RQ2, we looked at how different categories have evolved within the Ruby ecosystem. We found that categories like \textit{Application Quality and Security} have consistently generated a high volume of questions from developers since $2012$, highlighting some long-standing challenges. By addressing these recurring issues, developers and Ruby maintainers can focus on improving frameworks and libraries in ways that make Ruby development smoother and more efficient.

In RQ2, we found a decreasing trend in many of the topics on SO over time. 
This observation aligns with the findings from SO’s annual developer surveys, which reflect a similar decline in the popularity of Ruby and Ruby on Rails. For instance, in $2013$ and $2014$, Ruby was ranked as the $10th$ most popular technology \cite{StackOverflow2015}, but has a continuous decline in popularity through the following years \cite{StackOverflow2019, StackOverflow2020, StackOverflow2024}.

This decline could be related to several factors, including Ruby’s rich features and dynamic nature that make optimization challenging \cite{chevalier2023evaluating} and performance issues \cite{klochkov2021improving}.
However, despite its decline in usage, Ruby remains one of the top-paying technologies, ranked as the $5th$ highest-paying technology in the $2024$ SO survey \cite{StackOverflow2024, stackoverflowdevelopersurvey2023_2023}.
Therefore, understanding the challenges of Ruby developers paves the path to help Ruby developers' communities.

The difference we found in RQ3 and RQ4 results of SO vs. survey data can be attributed to several key factors that highlight the inherent differences in how both data sources measure difficulty. 
Survey data is subjective, relying on developers’ perceptions that are influenced by personal experiences and prior knowledge among others. In contrast, SO data is based on objective, measurable metrics that capture the broader activity on the platform. 
As a result, comparing the two becomes challenging, as they measure fundamentally different aspects of the developers' experiences.

Developers' perceptions of difficulty are shaped by a variety of personal and contextual factors that are not reflected on SO metrics, such as their learning style and prior knowledge in a particular field. 
Similarly, the time it takes for a question to receive an accepted answer depends on various external factors, such as how long it takes for other users to see the question and provide a response. 
Thus, the time-to-resolution might not necessarily reflect how difficult a developer finds a problem.
So, the current SO metrics may not fully represent the complexity or perceived difficulty of a given problem. There may be room for improvement in the metrics used to assess difficulty on SO.

\textit{Implications for researchers and Ruby maintainers.}
RQ1 and RQ2 and survey results reveal that some topics such as \textit{Software Architecture and Performance}, \textit{Application Quality and Security} or \textit{Development Environment and Infrastructure}, are challenging, even for experienced developers. This indicates opportunities for researchers to understand the issues related to these topics, and develop new tools and techniques to help Ruby developers. Additionally, it provides insights for Ruby maintainers to address the challenges faced by Ruby developers in the identified topics and support the Ruby developers' community more in these areas. 

\textit{Implications for researchers.}
The findings from our research reveal a significant difference between developers' survey-based perceptions of difficulty and time to resolve the issues and the objective metrics derived from SO. As discussed earlier, several factors can affect the perceived difficulty and time to resolve an issue by developers, and similarly on SO. Though this might be only related to Ruby developers, a direction for researchers is investigating and developing new metrics to capture the difficulty level of topics in online Q\&A platforms, especially on SO. 

\subsection{Threats to Validity}

\textbf{Internal Validity:}
Our dataset includes only SO questions, which may not capture all Ruby-related challenges. While SO is a common data source in the literature \cite{haque2020challenges, ahmed2018concurrency, bagherzadeh2019going, yang2016security, cummaudo2020interpreting, ahasanuzzaman2018classifying}, we validated our findings with a developer survey. To ensure quality, we filtered participants, recruited a large pool, used attention-check questions, and manually reviewed responses. However, we acknowledge that our Prolific-based sample may not fully represent the broader Ruby developer population.

\textbf{External Validity:} 
Our research only investigates the challenges of Ruby developers and may not generalize to other programming languages.

\textbf{Construct Validity:} We ensured accurate Ruby-related post identification by following established tag selection methods \cite{yang2023understanding, ahmed2018concurrency, yang2016security, bagherzadeh2019going, haque2020challenges, wen2021empirical}, and conducting multiple rounds of manual analysis, supported by high inter-rater reliability. To mitigate threats in topic modeling, we used the elbow method and manual evaluation to determine the optimal number of clusters. We manually reviewed the taxonomy groupings and labels to minimize bias, following methods from prior work \cite{shah2023using}, incorporating the top $10$ relevant keywords from BERTopic, and cross-referencing these keywords with sample posts for consistency. In our statistical analysis, we addressed multicollinearity using VIF scores to ensure model validity.

\textbf{Conclusion Validity:}
We ensured statistical validity through comprehensive validation, including $chi$-squared tests, regression model evaluation, Mann-Kendall trend analysis at $95\%$ confidence level, and correlation analysis. We normalized percentages when comparing categories and standardized both survey and SO metrics for consistent scales in our regression analysis.

\section{Conclusion and Future Work}

Our study aimed to uncover the challenges Ruby developers face by analyzing posts from Stack Overflow, supported by a survey of Ruby developers. We created a comprehensive taxonomy of $35$ topics, grouped into six main categories. 
Future research could expand the analysis beyond Stack Overflow to platforms like GitHub or Reddit for a more comprehensive view of Ruby developer challenges. Longitudinal studies could explore how these challenges impact developer productivity and influence the shift to other languages. 
Our results showed a misalignment between SO and survey data, reflecting the need to develop other metrics to capture difficulty levels. 
Finally, examining how emerging technologies like AI can be integrated to help Ruby developers, could be a line of research.

\begin{acks}
This research is supported by a grant from the Natural Sciences and Engineering Research Council of Canada RGPIN-2019-05175.
\end{acks}

\bibliographystyle{ACM-Reference-Format}
\bibliography{unveiling_ruby}

\end{document}